\documentclass[twocolumn,aps,floatfix,showpacs,prb,superscriptaddress]{revtex4}
\usepackage{epsfig}
\usepackage{graphicx}
\usepackage{amsmath}
\usepackage{amssymb}
\usepackage{amsfonts}
\usepackage{tabularx}
\usepackage{color}
\usepackage{dcolumn}
\usepackage{bm}
\usepackage{floatflt}
\usepackage{setspace}

\newcommand{\urusi}  {URu$_2$Si$_2$}

\begin{document}

\author{P.\,M.\ Oppeneer}
\affiliation{Department of  Physics and Astronomy, Uppsala University, Box 516, S-75120 Uppsala, Sweden}
\author{S.\ Elgazzar}
\affiliation{Physics Department, Johannesburg University, P.O.\ Box 524, Auckland Park 2006, South Africa}
\author{J.\ Rusz}
\affiliation{Department of  Physics and Astronomy, Uppsala University, Box 516, S-75120 Uppsala, Sweden}
\author{Q.\ Feng}
\affiliation{Department of  Physics and Astronomy, Uppsala University, Box 516, S-75120 Uppsala, Sweden}
\author{T.\ Durakiewicz}
\affiliation{Los Alamos National Laboratory, Condensed Matter and Thermal Physics Group, Los Alamos, NM 87545, USA}
\author{J.\,A.\ Mydosh}
\affiliation{Kamerlingh Onnes Laboratory, Leiden University, NL-2300 RA
Leiden, The Netherlands}

\title{Spin and orbital hybridization at specifically nested Fermi surfaces in  {\urusi}}

\date{\today}

\begin{abstract}
 The Fermi surface (FS) nesting properties of {\urusi} are analyzed with particular focus on their implication for the mysterious hidden order phase.
We show that there exist two Fermi surfaces that exhibit a strong nesting at the antiferromagnetic wavevector, $\boldsymbol{Q}_0$=(0,\,0,\,1). The corresponding energy dispersions fulfill the relation
$\epsilon_{1}(\boldsymbol{k})$=$- \epsilon_{2} (\boldsymbol{k}\pm \boldsymbol{Q}_0)$ at eight FS hotspot lines.
The spin-orbital characters of the involved $5f$ states are {\it distinct} 
($j_z$=$\pm$5/2 {\it vs.} $\pm$3/2) and hence the degenerate Dirac crossings
are symmetry protected in the nonmagnetic normal state.  
Dynamical symmetry breaking through an Ising-like spin and orbital excitation mode with  $\Delta j_z$=$\pm$1 
induces a hybridization of the two states, causing substantial FS  gapping.
Concomitant spin and orbital currents in the uranium planes give rise to a rotational symmetry breaking.
\end{abstract}

\pacs{71.20.-b, 71.27.+a, 74.70.Tx}

\maketitle

At  temperatures below $T_{\rm o}$=17.5\,K a mysterious hidden order (HO) phase develops
\cite{palstra85,maple86,schlabitz86}
 in the heavy-fermion uranium compound {\urusi}, the origin of which could not be definitely established
 despite intensive investigations.\cite{mydosh11}
The occurrence of the new, ordered phase below $T_{\rm o}$ is clearly witnessed by a sharp, second order phase transition appearing in the thermodynamic and transport quantities.
\cite{palstra85,schlabitz86,schoenes87,behnia05} 
The appearing new electronic order is not long-range ordered (dipolar) magnetism, although a small pressure of about 0.5 GPa suffices to stabilize long-range antiferromagnetic (AF) order.
\cite{amato04,amitsuka07} 
Recent experimental progress succeeded to reveal particular features of the HO state,
 thus providing a mosaic of pieces for unraveling the HO.\cite{wiebe07,santander09,schmidt10,aynajian10,bourdarot10,hassinger10,yoshida10,altarawneh11,dakovski11}
To explain the origin of the HO  a large number of sometimes exotic theories have been proposed over a period of more than twenty years\cite{santini94,ikeda98,chandra02,mineev05,varma06,elgazzar09,haule09,cricchio09,harima10,dubi11}
which however could not yet provided a complete understanding (see Ref.\ \onlinecite{mydosh11} for a survey of theories).

 A central question concerning the nature of the HO phase
  is which symmetry is spontaneously broken at the HO transition. A recent torque experiment performed on very small ($\mu$m-size) single crystals  measured the magnetic susceptibility in the basal $a$--$a$ plane of the tetragonal unit cell. \cite{okazaki11} Okazaki {\it et al.}\cite{okazaki11} observe rotational symmetry breaking, i.e., the off-diagonal susceptibility $\chi_{xy}$ is nonzero in the HO phase, unlike in the normal nonmagnetic phase above $T_{\rm o}$ where $\chi_{xy}$=0. 
 To explain the nonzero off-diagonal susceptibility several novel models\cite{thalmeier11,pepin11,fujimoto11}  for the HO phase have  been put forward recently. The non-vanishing off-diagonal susceptibility has been ascribed to a certain type of quadrupolar order,\cite{thalmeier11}  a modulated spin liquid,\cite{pepin11} and a spin nematic state. \cite{fujimoto11}

Apart from breaking of  $xy$-symmetry in the basal plane, it recently became clear that the lattice periodicity along the $c^{\star}$-axis in the Brillouin zone (BZ) is modified, too, in the HO phase.\cite{hassinger10,yoshida10}
The body-centered tetragonal (\textsc{bct}) unit cell of {\urusi} in the normal state becomes doubled 
in the HO state, and thus becomes simple tetragonal (\textsc{st}), consistent with a recent prediction.\cite{elgazzar09}
A complete understanding of the HO state obviously requires an explanation of why both these two symmetry breakings occur, something which has not yet emerged.

Here, we investigate the materials specific characteristics of the nested Fermi surfaces in {\urusi} and their consequences for the HO.
The existence of a FS instability had been predicted by density-functional theory (DFT) based calculations.\cite{elgazzar09} However, the importance of the occurring degenerate band-crossings in relation to the HO, and the full implications of FS nesting and the involved states have not yet been realized. 
Here we show that there exist two FS sheets in the paramagnetic \textsc{bct} phase that are strongly nested with nesting vector $\boldsymbol{Q}_0$=(0,\,0,\,1), and that the corresponding dispersions fulfill
$\epsilon_{1}(\boldsymbol{k})$=$- \epsilon_{2} (\boldsymbol{k}\pm \boldsymbol{Q}_0)$. 
A hybridization between the 
two crucial uranium $5f$ states can be created by a spin-orbital fluctuation with $\Delta j_z$=$\pm$1 which, as we discuss, has implications for the properties of {\urusi} in the HO.

\begin{figure}[tbh]
  \includegraphics[angle=90,width=0.75\linewidth]{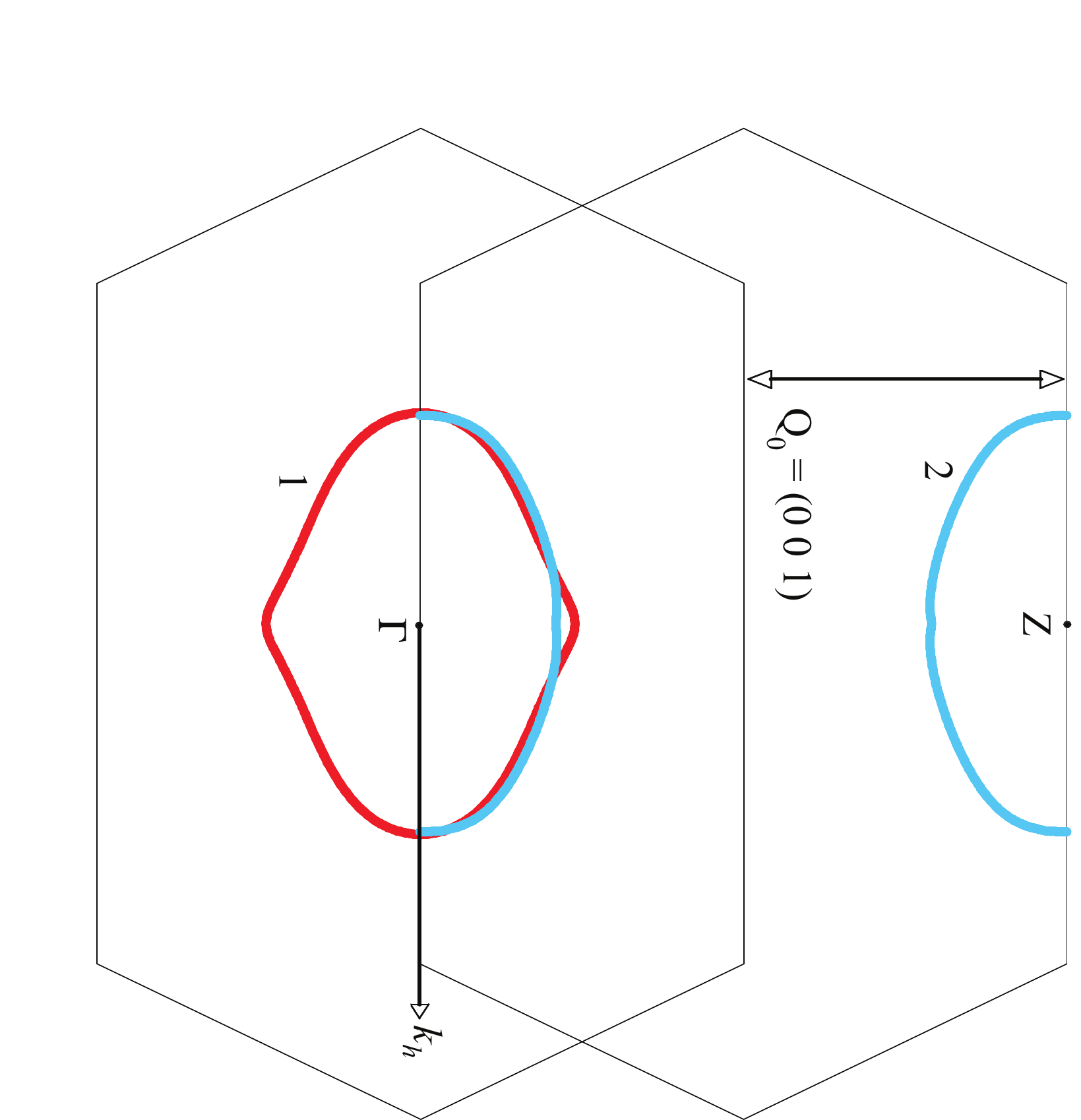}
  \vspace{-0.7cm}
  \caption{(Color online) Nesting of the two essential FS sheets of {\urusi},
  labeled No.\ 1 and 2, at $\boldsymbol{Q}_0$, illustrated for a cross-section of the \textsc{bct} BZ. The cross-section has been taken through two hotspot lines, along $k_h$$\approx$(0.5,\,0.255,\,0) and includes the $k_z$=(0,\,0,\,1) axis ($\Gamma$-Z).  The upper FS cross-section at Z has been shifted by $\boldsymbol{Q}_0$ to illustrate the nesting.
  \label{fig1}}
\end{figure}

\begin{figure}[tbh]
  \includegraphics[width=0.65\linewidth]{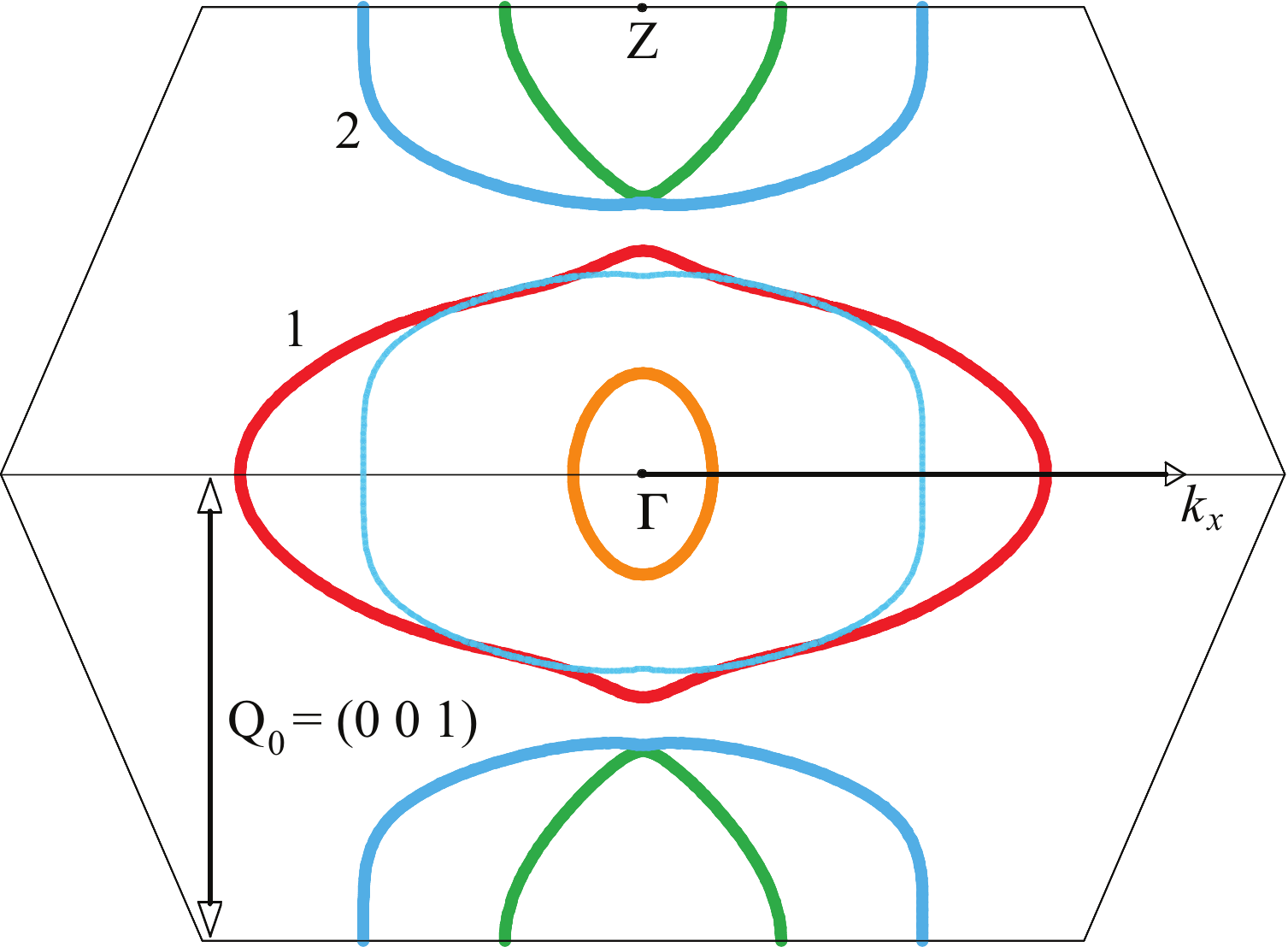}
  \vspace{-0.2cm}
  \caption{(Color online) Cross-section of the \textsc{bct} BZ illustrating an imperfect nesting of the two essential FS sheets No.\ 1 and 2 at $\boldsymbol{Q}_0$. FS sheet No.\ 2  has been shifted by $\boldsymbol{Q}_0$,  and is shown by the thin light grey (color online: light blue) curve, to illustrate the imperfect nesting. The cross-section is spanned by the $k_z$ and $k_x$ axes.
 The cross-sections of two other FS pockets are shown by the green and orange lines. 
  \label{fig2}}
\end{figure}

Several features of {\urusi} have recently become firmly established. 
Quantum oscillation studies\cite{ohkuni99,hassinger10,altarawneh11} have provided a consistent picture of the extremal FS orbits.
Earlier and recent  inelastic neutron scattering experiments\cite{broholm91,wiebe07,villaume08} revealed  two important FS nesting vectors,
$\boldsymbol{Q}_0$=(0,\,0,\,1), the antiferromagnetic or commensurate wavevector,
and $\boldsymbol{Q}_1$=(1$\pm$0.4,\,0,\,0), the incommensurate wavevector.
At these vectors, pronounced long-lived spin resonances have been detected in the HO state.\cite{bourdarot10} 
First-principles DFT calculations\cite{elgazzar09,oppeneer10} have predicted a FS in good agreement with the quantum oscillation studies\cite{ohkuni99,hassinger10,altarawneh11} and also consistent with recent high-energy angle-resolved photoemission spectroscopy.\cite{kawasaki11} 
Evidently, this FS is the appropriate starting point for materials specific studies of the HO.
Our focus here is on the AF commensurate nesting vector $\boldsymbol{Q}_0$, which is half a $\Gamma$--$\Gamma$ reciprocal lattice distance and is favorable for the appearance of long-range AF order at modest pressure of 0.5\,GPa and  AF spin excitations in the HO phase.\cite{elgazzar09,oppeneer10}

Figure \ref{fig1} shows the nesting of the two involved FS sheets at $\boldsymbol{Q}_0$.
The DFT calculations identified--in the \textsc{bct} normal state--a $\Gamma$-centered closed FS sheet with an approximate ellipsoidal shape as well as a closed Z-centered FS sheet; these are labeled FS sheet Nos.\ 1 and 2. Apart from these two sheets there are three other closed FS sheets, a very small $\Gamma$-centered ellipsoid,  a Z-centered ellipsoid, and an ellipsoid centered at the X point in the \textsc{bct} BZ. These are not shown in Fig.\ \ref{fig1} as they do not exhibit pronounced nesting properties.
The $\Gamma$-centered sheet (dark grey, red in color online) and Z-centered sheet (light grey, blue in color online) are strongly nested with nesting vector $\boldsymbol{Q}_0$. On these two FS sheets there are eight hotspot lines,\cite{elgazzar09} where the two FS sheets intersect when shifted by $\boldsymbol{Q}_0$. Figure \ref{fig1} illustrates the degree of nesting at a cross-section of the BZ going approximately through two FS hotspot lines as well as the $\Gamma -$Z axis. The good degree of nesting of the two FS can be recognized when one overlays them (note that the lower half of FS No.\ 2 is shown unshifted). On the hotspot lines the nesting is in fact perfect, but since the hotspot lines are not completely straight there is a small deviation in the depicted cross-section. Also, the $\Gamma$-centered FS sheet has a pointed part close to -Z/2 and at Z/2. This ``nipple" prevents a complete nesting of the two FSs.

Away from the hotspot lines the nesting is not equally good. This is illustrated in Fig.\ \ref{fig2} where we show the FS cross-sections in another plane in the \textsc{BZ}, here the $k_z - k_x$ plane. In this figure we also show two of the three other existing pockets, the $\Gamma$- and Z- centered ellipsoids.  A good nesting of the two essential FS sheets 1 and 2  is present in the reciprocal space regions halfway between $\Gamma$ and Z and  $\Gamma$ and -Z, again with the exception of the nipple at Z/2 and -Z/2. The nesting becomes particularly poor in the $k_x$ direction. A similar \textit{imperfect} nesting occurs along other planes cutting the BZ away from the hotspot lines, for example the plane spanned by $k_z$ and the vector (1,\,1,\,0) (not shown here).

In the energy bandstructure there are two bands which, when folded through the AF wavevector, exhibit an accidental degenerate crossing.\cite{elgazzar09} At the hotspot lines this degenerate band crossing leads to a Dirac point occurring precisely at the Fermi level,\cite{oppeneer10} where the band dispersions are approximately linear in the vicinity of the crossing point. Away from the hotspot lines
the degenerate band crossing point falls either above (along the $\Gamma$--(1,\,1,\,0)  direction) or below (along the $\Gamma$--(1,\,0,\,0) direction) the Fermi level. 
The relation between the energy dispersions of the two nested Fermi surfaces has not yet been given. An inspection of the bands shows that it is: $\epsilon_{1}(\boldsymbol{k})$=$- \epsilon_{2} (\boldsymbol{k}\pm \boldsymbol{Q}_0)$. Nesting conditions of similar form, $\epsilon_1(\boldsymbol{k})$=$- \epsilon_1 (\boldsymbol{k} + \boldsymbol{Q})$, have drawn attention previously in the context of spin-density waves, orbital antiferromagnetic and spin  nematic order for systems with half-filled bands.\cite{nersesyan91,schulz89}  
Recently, it was applied to {\urusi} to propose spin nematic order.
\cite{fujimoto11}

\begin{figure}[tb!]
  \includegraphics[angle=0,width=0.45\textwidth]{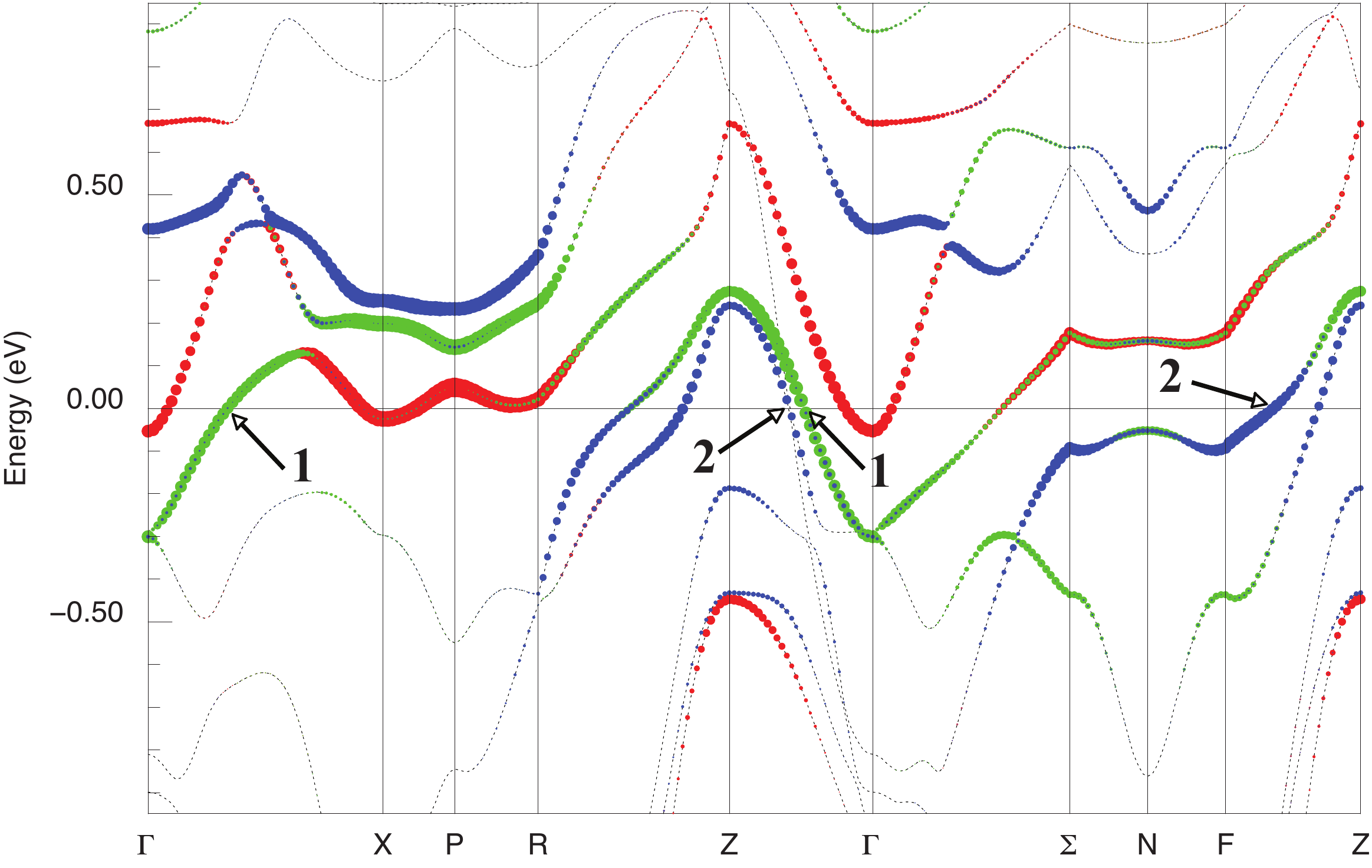}
  \caption{(Color online) 
  Band characters of the energy dispersions of  {\urusi} in the \textsc{bct} BZ. Green symbols illustrate the U $5f_{5/2}$ $j_z$=$\pm$5/2 character, blue symbols the $j_z$=$\pm$3/2 character, and red symbols the $j_z$=$\pm$1/2 character 
(for the used symmetry points, see Ref.\ \onlinecite{oppeneer10}).
   \label{fig3}}
\end{figure}

The orbital character of the band states of the nested FS sheets is an important aspect. As spin-orbit interaction is strong in {\urusi}, spin and orbital angular momenta are not good quantum numbers; the total angular moment $j$ is appropriate here. All uranium $5f$ states in the vicinity of $E_F$ stem from the $j$=5/2 manifold.\cite{oppeneer10} The $j_z$ character of the bands near the Fermi energy is shown in Fig.\ \ref{fig3}. The relativistic band structure  calculations have been performed with the Kohn-Sham-Dirac DFT formalism in the local density approximation (LDA).\cite{oppeneer10}
The thickness of the symbols illustrates the amount of $j_z$ spin-orbital character. This plot highlights that the 
U $5f$-states of the $\Gamma$-centered and X-centered ellipsoids have strongly $j_z$=$\pm$1/2 spin-orbital character,  the $\Gamma$-centered FS No.\ 1
has $j_z$=$\pm$5/2 character, whereas the large Z-centered FS sheet No.\ 2 has dominantly $j_z$=$\pm$3/2 character.\cite{note}
The U $5f$ states in the two bands creating the degenerate crossing have thus distinct spin-orbital characters, differing by $\Delta j_z$=$\pm$1.

The unequal spin-orbital characters of the two states has pertinent consequences for how they can interact, as these states of different symmetry cannot hybridize in the normal state, and therefore cannot lift the degeneracy.
This is analogous to the Dirac crossing found in topological insulators, where the degenerate crossing of $s_z$=+1/2  and  $s_z$=$-1/2$  bands is symmetry protected by time-reversal symmetry.  In {\urusi} a coupling between the two states requires a spontaneous symmetry-breaking with $\Delta j_z$$\ne$0. This can be achieved by a longitudinal moment excitation with $j_z$$\ne$0 which we propose to be the responsible quasiparticle acting as the coupling mediator between electrons in the two spin-orbital states. Such an AF mode at $\boldsymbol{Q}_0$ guarantees the required conservation of angular momentum $j_z$ and linear momentum, $\boldsymbol{k} - \boldsymbol{k}'$=$\pm \boldsymbol{Q}_0$.  Moment excitations with $\Delta j_z$=1 and $\Delta j_z$=-1 are equally possible, as the nonmagnetic states in the normal phase contain equal amounts of $+j_z$ and $-j_z$ character.  The required coupling of the two states over the AF vector $\boldsymbol{Q}_0$ is provided by staggered moment excitations on the U atoms in the basal plane and at the \textsc{bct} vector, i.e., breaking \textsc{bct} lattice symmetry. 

An alternative spontaneous symmetry breaking would be a static spin density wave, that is, static AF order at $\boldsymbol{Q}_0$, which indeed is stabilized in {\urusi} at modest pressures. DFT calculations\cite{elgazzar09} predicted a very small energy difference ($\sim$14\,K) between the normal and AF ordered states. The AF moment excitations are expected to occur at even lower energies.

The essential physics of the band hybridization can be captured in a simple model.
Note that only dispersive bands are involved, not a hybridization between a localized $f$ level and a conduction band.
For the two bands only four spin-orbital $5f$ states are relevant, $j_z$=$\pm$3/2 and $\pm$5/2. Denoting the two U atoms as U$_A$ (basal plane) and U$_B$ (at \textsc{bct} position), the degenerate states in the folded BZ ($\boldsymbol{k'}\pm\boldsymbol{Q}_0$$\rightarrow$$\boldsymbol{k}$) at the band crossing can be represented as
$| 1 \rangle_{\alpha; \pm}$= $|1\rangle_{\alpha ,A} \pm |1\rangle_{\alpha ,B}$ and
$| 2 \rangle_{\beta; \pm}$= $|2\rangle_{\beta ,A} \pm |2\rangle_{\beta ,B}$, with 
$| 1 \rangle_{\alpha =\pm, A}$= $|5/2\rangle_{A} \pm |$-$5/2\rangle_{A}$ and 
$| 2 \rangle_{\beta =\pm, A}$= $|3/2\rangle_{A} \pm |$-$3/2\rangle_{A}$ and analogous for atom $B$.
 A coherent spin-orbital fluctuation by $\Delta j_z$=1 on U$_A$ and $\Delta j_z$=-1 on U$_B$ leads to states $|1'\rangle_{\alpha; \pm}$, $|2 '\rangle_{\beta; \pm}$, for example,   $|1' \rangle \propto |5/2 \rangle_A+  |$-$3/2 \rangle_A + |3/2 \rangle_B +|$-$5/2\rangle_B$ and $|2' \rangle \propto |5/2 \rangle_A+  |$-$3/2 \rangle_A + |3/2 \rangle_B +|$-$5/2\rangle_B$, allowing hybridization. A spin-orbital excitation with $\Delta j_z$=-1 on U$_A$ and $\Delta j_z$=1 on U$_B$ gives the same, but with the role of U$_A$ and U$_B$ interchanged.
The proposed quasiparticle that facilitates hybridization of the two band states 
is henceforth a dynamical mode of $\boldsymbol{Q}_0$ longitudinal AF excitations.  
Inelastic neutron scattering experiments have detected an intensive, long-lived magnetic resonance at $\boldsymbol{Q}_0$, \cite{broholm91,villaume08,bourdarot10} which was proposed to be the dynamical driving mechanism of the HO transition.\cite{elgazzar09,oppeneer10}
For completeness we mention that there exists also the incommensurate wavevector $\boldsymbol{Q}_1$ at which magnetic excitations also occur
\cite{broholm91,wiebe07} which alternatively have been proposed\cite{balatsky09,schmidt10,dubi11} as the mechanism responsible for the HO.
We find however that folding of the band dispersions at $\boldsymbol{Q}_0$ leads to a much more pronounced FS gapping.\cite{oppeneer10}

The dynamical AF mode satisfies the $\Delta j_z$=$\pm1$ condition and additionally explains other peculiar properties of {\urusi}. Early inelastic neutron experiments found a magnetic longitudinal excitation with an unexpected large matrix element of $g \mu_B | j_z  | \approx 1.2 - 2.2 $ $\mu_B$,\cite{broholm87,broholm91}
which is consistent with the $\Delta j_z$ required for interaction of the band states. 
Another property of {\urusi} which was not yet explained is the peculiarity that its magnetic excitations are longitudinal, i.e., Ising-like along the $c$-axis. In contrast, common low-energy excitations in conventional magnetic materials are transversal moment fluctuations. This unusual  feature of the HO phase of {\urusi} follows here from the orbital characters of the $j_z$-bands. 

Right at the Dirac crossings there exists good particle-hole symmetry. Anomalous particle-hole pairing at bands  with nesting condition $\epsilon_1(\boldsymbol{k})$= $- \epsilon_1 (\boldsymbol{k} + \boldsymbol{Q})$ --which can give rise to orbital AF or spin nematic orders\cite{schulz89,nersesyan91,fujimoto11} --can therefore appear at the hotspot lines. However, this would be for bands with {\it identical} character and the situation in {\urusi} is distinct.
First, the unequal spin-orbital character of the involved bands should be taken into account rendering the Dirac crossing point symmetry protected. Second, the particle-hole pairing gap opened at the hotspots would be symmetric around the Fermi energy, but STS experiments detected a gap that is not symmetric around $E_F$.\cite{aynajian10} Furthermore, particle-hole symmetry is not preserved at $k$-points not located directly on the hotspot lines. Accordingly, the particle-hole pairing gap opening at the FS would then be restricted to the immediate vicinity of the hotspot lines which implies only a modest gapping of the total FS. Experiments however find a substantial removal of about 40\% of the FS in the HO.\cite{maple86} 
For {\urusi} the AF excitation is a prerequisite for the hybridization. 
Moreover, the AF mode provides hybridization of $\boldsymbol{Q}_0$-folded states $|1\rangle$ and $|2\rangle$ over a wide $k$-space region and induces a much larger FS gapping, in which degenerate crossings above and below the Fermi level become gapped, too.\cite{oppeneer10} Upon AF excitation anomalous particle-hole pairing at the hotspots becomes possible which gives rise to staggered spin and charge currents\cite{schulz89,nersesyan91} around U atoms in the tetragonal planes. As pointed out previously
\cite{nersesyan91} (for the situation without AF mode),
 such occurring static currents cause rotational symmetry breaking of the underlying tetragonal lattice and the spin currents break SU(2) spin rotational symmetry.
Static staggered charge currents in the plaquette of U atoms would be detectable with neutron scattering or $\mu$SR, but these have never been observed in {\urusi}. 

A consequence of the dynamical spin-orbital mode is the appearance of dynamical spin and charge currents circulating around the U atoms in the plaquettes, exhibiting antiparallel rotation helicity for U atoms connected through the \textsc{bct} vector.
To analyze the consequences of the circulating currents we consider the two band states forming the nested FS as a minimal yet sufficient model. Circulating charge currents $j_{\alpha}=q p^{\alpha}$ around atom U$_A$ give rise to a non-vanishing off-diagonal conductance, 
$\sigma_{xy} \propto 
i q^2\sum_{i \ne j} (f_{j} - f_{i}) \mathcal{P} \frac{1}{(\epsilon_{j} - \epsilon_{i})^2}   p^x_{ij} p^y_{ji}$,
 with $i,j =|1'\rangle_A$ or $|2'\rangle_A$, 
 $q$ the charge, $f$ the Fermi function, and $\vec{p}$ the momentum operator. 
This gives $p^x_{1'2'} p^y_{2'1'}$$\ne$0.
Likewise, circulating spin currents $j_{\alpha}=\vec{s} p^{\alpha}$ lead to a non-vanishing off-diagonal spin conductance. Considering its component in the basal plane, $\propto 
i \sum_{i \ne j} (f_{j} - f_{i}) \mathcal{P} \frac{1}{(\epsilon_{j} - \epsilon_{i})^2}   (s^xp^x)_{ij} (s^y p^y)_{ji}$, 
and using that $s^x_{ii}$=$s^y_{ii}$=0 (but $p^x_{ii}$=$\frac{m}{\hbar} \partial \epsilon_i/ \partial k_x$$ \ne$0)
one obtains $s^x_{1'2'} s^y_{2'1'}$$\ne$0.
 As a result
 the off-diagonal spin susceptibility $\chi_{xy} \propto -\sum_{i \ne j}
 (f_{j} - f_{i}) \mathcal{P} \frac{1}{(\epsilon_{j} - \epsilon_{i})}  s^x_{ij} s^y_{ji}$, which contains terms $s^x_{1'2'} s^y_{2'1'} $ is nonzero.
%
When the charge current direction reverses, $\sigma_{xy}$ changes sign (as $\sigma_{yx}$=$ - \sigma_{xy}$) and thus the averaged off-diagonal conductance vanishes when averaged over a sufficiently long time. The averaged off-diagonal susceptibility however does not vanishes, as $\chi_{xy}$=$\chi_{yx}$. Hence, due to the occurrence of circulating charge and spin currents, $\chi_{xy}$$\ne$0, giving a rotational symmetry breaking consistent with the nonzero susceptibility observed in recent torque measurements.\cite{okazaki11}

The dynamic aspect of the AF resonance is an essential difference between the HO and the long-range ordered AF phase. In the latter phase the moments are static and no oscillating currents appear in the U planes. The only breaking of the lattice symmetry is the doubling of the unit cell, giving a modulation of the $c^{\star}$-axis periodicity. In the HO phase conversely the interaction of bands folded over the AF  $\boldsymbol{Q}_0$ vector leads to the observed\cite{hassinger10,yoshida10} modulation of the lattice periodicity along the $c^{\star}$-axis, and in addition the circulating currents lead to breaking of rotational symmetry in the basal planes.\cite{okazaki11}

To summarize, specific FS nesting conditions exist in  {\urusi} which are pertinent to its HO. 
An Ising-like excitation mode induces an interaction of two FS states connected by the $\boldsymbol{Q}_0$ vector, leading to hybridization, FS gapping, and lattice symmetry breaking in accord with recent experiments.
This emphasizes that HO is a momentum space ordering phenomenon, rather than a property of a localized $f$ configuration.

We thank  S.\ Fujimoto for a helpful discussion.
This work was supported by the Swedish Reseach Council (VR) and Swedish National Infrastructure for Computing (SNIC).


\begin{thebibliography}{99}
\bibitem{note}
Note that the $j_z$ character  changes in the poorly nested lobes of FS No.\ 1, which do not participate in the gapping.
\end{thebibliography}
\end{document}